\documentclass[twocolumn,prd,aps,showpacs,superscriptaddress]{revtex4}
\usepackage{natbib}
\usepackage{amsmath}
\usepackage{verbatim}
\usepackage{graphicx}
\usepackage[usenames]{color}
\usepackage{psfrag}

\usepackage{epstopdf}
\usepackage{amsbsy,amssymb,amsmath}
\usepackage{rotating}

\usepackage{psfrag}
\usepackage[ps2pdf,colorlinks,bookmarks]{hyperref}
\definecolor{linkblue}{rgb}{0,0,0.8}
\definecolor{linkgreen}{rgb}{0,0.5,0}
\hypersetup{pdfpagemode=None, pdfstartview=FitH, linkcolor=linkblue, %
            citecolor=linkgreen, urlcolor=linkblue}

\def\beq{\begin{equation}}
\def\eeq{\end{equation}}
\def\bea{\setlength\arraycolsep{1.4pt}\begin{eqnarray}}
\def\eea{\end{eqnarray}}
\def\bit{\begin{itemize}}
\def\eit{\end{itemize}}

\begin{document}

\title{Tilted Physics: A Cosmologically Dipole-Modulated Sky}

\author{Adam Moss} \email{adammoss@phas.ubc.ca}
\affiliation{Department of Physics \& Astronomy\\
University of British Columbia, Vancouver, BC, V6T 1Z1  Canada}

\author{Douglas Scott} \email{dscott@phas.ubc.ca}
\affiliation{Department of Physics \& Astronomy\\
University of British Columbia, Vancouver, BC, V6T 1Z1  Canada}

\author{James P. Zibin} \email{zibin@phas.ubc.ca}
\affiliation{Department of Physics \& Astronomy\\
University of British Columbia, Vancouver, BC, V6T 1Z1  Canada}

\author{Richard Battye} \email{rbattye@jb.man.ac.uk}
\affiliation{Jodrell Bank Center for Astrophysics, School of Physics and Astronomy, University of Manchester, Manchester, M13 9PL  UK}

\date{\today}

\begin{abstract}
Physical constants and cosmological parameters could vary with position. On the largest scales such variations would manifest themselves as gradients across our Hubble volume, leading to dipole-modulation of the cosmic microwave anisotropies.  This generically leads to a correlation between adjacent multipoles in the spherical harmonics expansion of the sky, a distinctive signal which should be searched for in future data sets.
\end{abstract}

\maketitle

{\em Introduction.---}%
Physical laws are described by a set of fundamental {\em dimensionless\/} constants. These include the electromagnetic fine structure constant, $\alpha_{\rm e} \equiv e^2/ (4 \pi \epsilon_0 \hbar c)  \simeq 1/137$, and the much weaker gravitational fine structure constant, $\alpha_{\rm G} \equiv G m_{\rm p}^2/ (\hbar c) \simeq 6 \times 10^{-39}$ (here $m_{\rm p}$ is the proton mass; see e.g.\ \cite{Moss:2010ka} and references therein). The standard model of cosmology (SMC) (see e.g.\ \cite{scott06}) has an additional set of adjustable parameters, some of which may derive from the first group. However, we cannot yet calculate these parameters from the known physical laws, and indeed it is not clear that we will ever be able to do so in practice.

It is possible that any of these constants or parameters could vary in spacetime. Such a discovery would point to physics beyond the standard paradigms; for example, a spatial variation could indicate that separate patches of the Universe have different realizations of physics, possibly related to the string landscape. It is important to constrain such variations using available cosmological data, since one might expect effects to show up on the largest observable scales.

There has been a significant amount of activity investigating the potential time variation of $\alpha_{\rm e}$ (see Ref.~\cite{Uzan:2010pm} for a review).  Recently, studies of high redshift quasars have led to a claim of evidence for a spatial gradient (or ``tilt'') in $\alpha_{\rm e}$~\cite{Webb:2010hc}.  Specifically, a dipole anisotropy of the form $\Delta \alpha_{\rm e}/\alpha_{\rm e} = B (t_0 - t(z)) \cos\theta + m$ was fitted to quasar spectra, where $B$ and $m$ are constants, $t_0 - t(z)$ is the look-back time (as a function of redshift $z$), and $\theta$ is the angle with respect to the dipole orientation.  It was claimed that $B > 0$ at $\sim$4$\sigma$ significance, although this result is yet to be tested independently.  

An equally surprising claim made within the last few years has been that there is a very large-scale ``cosmic flow''~\cite{2008ApJ...686L..49K,2009MNRAS.392..743W}. Such a flow would be in conflict with predictions of the SMC, but could be produced by a large-scale isocurvature mode.
  
A gradient in $\alpha_{\rm e}$ or any other parameter could be observed (or constrained) at an early epoch using the CMB. For some of the parameters (e.g.\ $\alpha_{\rm e}$), the imprint on the CMB arises from a modification of recombination physics, giving an epoch of last scattering 
which is anisotropic.  For other parameters (e.g.\ the primordial amplitude), a gradient would be directly reflected in the properties of the primary CMB anisotropies.  For another class of parameters (e.g.\ the density parameters), a gradient would correspond to a long-wavelength isocurvature mode.  In this Letter we describe how one would search for this effect using the higher order multipoles of the CMB covariance matrix.

Several previous studies have discussed the breaking of statistical isotropy in the primordial fluctuations (e.g.\ \cite{Gordon:2005ai}) and others have looked at the effects and observability of dipole or quadrupole multiplicative factors applied to the temperature field on the sky (e.g.\ \cite{Zheng:2010ty,Hanson:2009gu}).  Although these ideas are related to ours, here we are considering spatial variations in the {\it underlying parameters\/} themselves.  In Ref.~\cite{Donoghue:2004gu}, the effect of spatial variation of cosmological parameters on the CMB was studied, although the covariance matrix was not considered.

\begin{figure}
\centering \mbox{\resizebox{0.32\textwidth}{!}{\includegraphics[angle=0]{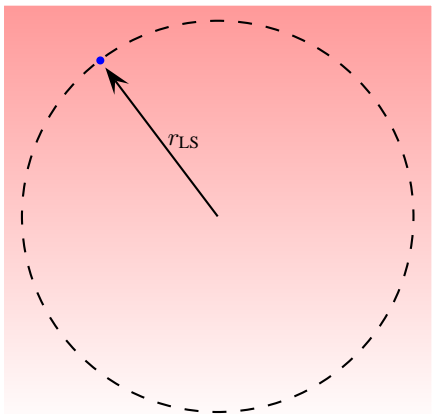}}} 
\caption{\label{fig:gradient} Spatial gradient in a cosmological parameter (represented by the colour gradient), which is locally negligible at last scattering, but important over the free-streaming length scale.  The filled circle shows (to scale) the comoving horizon size at last scattering.}
\end{figure}

{\em Dipole modulation of physics.---}%
To make the problem tractable within standard perturbation theory, we assume that there is little spatial variation of each parameter inside the Hubble radius at the time of recombination, $t_{\rm rec}$.  This means that insofar as the local physics is concerned, the gradient can be considered a first-order perturbation, and hence we can use the Friedmann-Lema\^itre-Robertson-Walker background to evolve the matter perturbations as usual up to $t_{\rm rec}$, with the parameter value(s) depending on the location on the last scattering surface (LSS).  Today, however, we are observing photons free streaming from much greater distances, so the spatial gradient may become noticable over these scales (see Fig.~\ref{fig:gradient}).

The spatial variation could be of arbitrary complexity, but here we restrict our attention to a linear gradient across our Hubble volume. This is motivated partly by the suggestions of a gradient in $\alpha_{\rm e}$ and by the flow measurements, but can also be justified in that any residual inhomogeneity from before inflation would be expected to have been stretched until it was essentially linear inside our Hubble volume.  Therefore we characterize the anisotropy in parameter $X$ at $t_{\rm rec}$ by \beq X({\hat{\bf{n}}}) =  X_0 \left(1  + \Delta X/X_0 \right)
 \cos \theta \,,
\eeq where ${\hat{\bf{n}}}$ is the direction to the LSS and $X_0$ is the monopole value.  We find that such a dipole induces correlations between the $\ell$ and $\ell+1$ multipole modes of the CMB covariance matrix, with an amplitude and spectrum depending on $X$. This is a generic feature of a dipole anisotropy---our motion with respect to the CMB rest frame \cite{Kosowsky:2010jm,Amendola:2010ty} or a Hubble-scale homogeneous magnetic field \cite{DurrerKY98,BernuiHR08} also induce similar correlations. 

In the following we fix the monopoles $X_0$ to their respective {\it WMAP\/} 7-year best-fit values~\cite{Komatsu:2010fb}.  As well as $\alpha_{\rm e}$, we consider the baryon matter fraction $f_{\rm b} = \Omega_{\rm b}/\Omega_{\rm m}$ at the LSS, and the amplitude and spectral index of the primordial fluctuations, $A_{\rm s}$ and $n_{\rm s}$.  A wider list could be considered, such as: $m_{\rm p}/m_{\rm e}$; other particle couplings, mass ratios, and mixing angles; helium abundance; or effective number of neutrino species.  However, our method assumes that the gradient affects only the generation of {\em primary} anisotropies (i.e.\ anisotropies generated at last scattering), so it should be generalized for parameters such as the redshift to reionization or tensor contribution, which effect anisotropies along the line of sight.


{\em Isocurvature mode.---}%
Some variations may lead to an additional $\ell=1$ mode.  In the case of motion with respect to the CMB frame, any difference between the dipole deduced from aberration and the directly measured CMB dipole can be ascribed to an isocurvature mode.   In the same way, gradients in parameters like $f_{\rm b}$ will correspond to gradients in the relative densities of matter components, i.e.\ to isocurvature modes, and hence will give rise to an ``intrinsic'' dipole on the sky.  On the other hand, modulation of other parameters, such as $A_{\rm s}$, will clearly not correspond to isocurvature modes.  When an isocurvature mode is present, the evolution of the mode should properly be examined to late times, to capture any line-of-sight effects that our approach would miss.  Therefore, we present the higher order correlations for $f_{\rm b}$ with the caveat that the connection with isocurvature models is yet to be fully explored.


{\em CMB covariance matrix.---}%
The CMB temperature anisotropies observed today are described by $T({{\hat{\bf{n}}}}) = T_0 \left[1+\Phi({{\hat{\bf{n}}}}) \right]$, where the $T_0$ is the monopole.
The fluctuations are typically expanded in terms of the spherical harmonics $Y_{\ell m }({{\hat{\bf{n}}}})$ as \beq \Phi({{\hat{\bf{n}}}})=\sum_{\ell=1}^{\infty}
 \sum_{m=-\ell}^{\ell} a_{\ell m} Y_{\ell m }({{\hat{\bf{n}}}}) \,.
\eeq In the approximation that the anisotropies are primary, we can Taylor expand the temperature field in terms of the unmodulated (statistically isotropic) field $\Phi^{\rm i}({{\hat{\bf{n}}}})$ and parameter $X$ to give \beq \Phi({{\hat{\bf{n}}}}) = \Phi^{\rm i}({{\hat{\bf{n}}}})
 + \Delta X \frac{d \Phi^{\rm i}({{\hat{\bf{n}}}}) }{dX}
 \sqrt{\frac{4\pi}{3}} Y_{10} ({{\hat{\bf{n}}}})\,,
\eeq where we have oriented the dipole along the polar direction.  The harmonic modes are then given by \beq a_{\ell m} = a_{\ell m}^{\rm i} + \Delta X 
 \sum_{l'} \frac{d a_{\ell' m}^{\rm i} }{dX}  \xi_{\ell m \ell' m}\,,
\eeq with the coupling coefficients  
\begin{eqnarray}
\xi_{\ell m \ell' m'} &=&    \delta_{\ell',\ell+1}\delta_{m m'}
 \sqrt{\frac{(\ell+1)^2 -m^2}{(2\ell+1)(2\ell+3)}} \\
 \nonumber  &+&   \delta_{\ell',\ell-1}\delta_{m m'}
 \sqrt{\frac{(\ell^2-m^2)}{(2\ell-1)(2\ell+1)}}  \,.
\end{eqnarray}
From this one can compute the correlation matrix 
\begin{eqnarray} \label{eqn:covmat}
C_{\ell m \ell' m'} &\equiv&\langle a_{\ell m}^* a_{\ell' m'} \rangle
  \\ \nonumber
 &=& C_{\ell} \delta_{\ell \ell'}  \delta_{m m'} +  \frac{\Delta X}{2}
  \left[ \frac{dC_{\ell}}{dX} + \frac{dC_{\ell'}}{dX}  \right]
  \xi_{\ell m \ell' m'}\,,
\end{eqnarray}
to order $\Delta X$, where
 $ \langle a^{\rm i *}_{\ell m} a^{\rm i}_{\ell' m'} \rangle
 = C_{\ell} \delta_{\ell \ell'}  \delta_{m m'}  $.  The only non-isotropic
terms are the off-diagonal $\ell, \ell \pm 1$ modes---as expected, a dipole modulation couples multipoles with $\left|\Delta\ell\right|=1$. The derivative power spectra can easily be extracted from standard CMB codes; we used CAMB~\cite{Lewis:1999bs},  using the unlensed spectra.

The off-diagonal terms due to the aberration caused by our motion with respect to the CMB rest frame have a similar form, also coupling modes with $\left| \Delta \ell \right| =1$.  These are given by (see e.g.\ \cite{Kosowsky:2010jm}) \beq
 C_{\ell,m,\ell+1,m} \approx \beta (\ell+1)
  \left(C_{\ell+1}^{\rm } - C_{\ell}^{\rm }  \right)
  \xi_{\ell,m,\ell+1,m} \,,
\eeq where $\beta=v/c \approx 1.23 \times 10^{-3}$.  This is essentially a derivative of the $C_{\ell}$ with respect to $\ell$ itself, since $ C_{\ell+1}^{\rm } - C_{\ell}^{\rm } \approx d C_{\ell} / d \ell $. 


{\em Variable $\alpha_{\rm e}$ example.---}%
To choose a concrete example, let us focus on $\alpha_{\rm e}$. We recap the basic recombination equation, considering the 3-level hydrogen atom.  A full explanation of the symbols can be found in Ref.~\cite{Seager:1999bc}.  The rate equation  is 
 \begin{eqnarray}
H \frac{dx_{\rm p}}{d \ln z} &=&     \left[ x_{\rm e}x_{\rm p} n_{\rm H} 
 \alpha_{\rm H} - \beta_{\rm H} (1-x_{\rm p}) 
   {\rm e}^{-h\nu_{\alpha}/kT_{\rm M}} \right]  \\
  \nonumber  & \times& {1 + K_{\rm H} \Lambda_{\rm H} n_{\rm H}(1-x_{\rm p})
    \over 1+K_{\rm H} (\Lambda_{\rm H} + \beta_{\rm H})
     n_{\rm H} (1-x_{\rm p}) }\,,
\end{eqnarray}
where $x_{\rm p} = n_{\rm p} / n_{\rm H}$ is the ionization fraction, $\beta_{\rm H}$ the ionization rate, $\Lambda_{\rm H}$ the 2-photon rate, $K_{\rm H}=\lambda_{\alpha}^3/(8 \pi H(z))$ the redshift factor, and $\alpha_{\rm H}$ the recombination rate.

To adapt the RECFAST recombination code, we convert all energies and rates according to their  $\alpha_{\rm e}$ dependence.  The modifications are as follows: $\lambda_{\alpha} \rightarrow \lambda_{\alpha}
 \left(1+ \epsilon  \right)^{-2}$;
$K_{\rm H} \rightarrow K_{\rm H}
 \left(1+\epsilon  \right)^{-6}$;
$\alpha_{\rm H} \rightarrow \alpha_{\rm H}
 \left(1+\epsilon  \right)^5$;
$\beta_{\rm H} \,  \rightarrow   \beta_{\rm H} \,
 \left(1+\epsilon  \right)^5
 {\rm e}^{-\epsilon^2}$;
and $\Lambda_{\rm H} \rightarrow \Lambda_{\rm H}
 \left(1+\epsilon  \right)^8$,
 where $\epsilon = \Delta \alpha_{\rm e}/\alpha_{\rm e} \cos \theta$. We found the changes to the ionization history to be consistent with other work. 
An increase in $\alpha_{\rm e}$ leads to faster recombination, which 
is primarily due to the decrease in the ionization rate. 

\begin{figure}
\centering
\includegraphics[width=\columnwidth,angle=0]{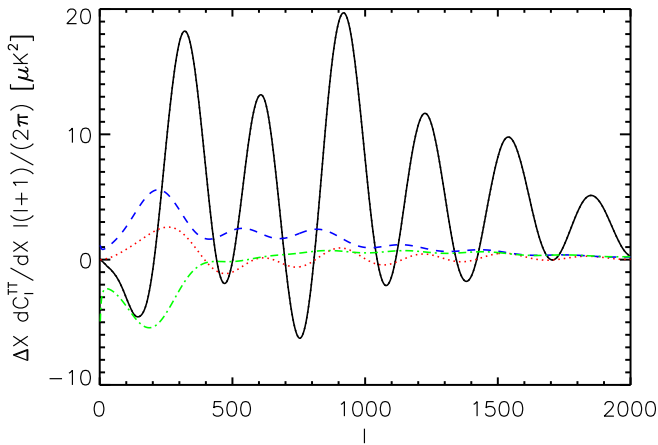} \caption{\label{fig:cls}  Derivative power spectra $\Delta X\, d C_\ell/dX$ with $ \Delta X/X_0 =  10^{-3}$  for $X = \alpha_{\rm e}$ (solid, black), $f_{\rm b}$ (dotted, red), $A_{\rm s}$ (dashed, blue), and $n_{\rm s}$ (dot-dash, green).}
\end{figure}

{\em Results.---}%
In Fig.~\ref{fig:cls} we show the derivative power spectra for several cosmological parameters with amplitude $\Delta X/X_0 = 10^{-3}$. The $C_{\ell}$ spectrum has a stronger relative dependence on  $\alpha_{\rm e}$ than other parameters,  giving larger off-diagonal correlations. We can see that the spectra have different shapes, and hence the dipole modulation of various parameters could in principle be distinguished.

Note that the spectrum for $X = n_{\rm s}$ depends on the chosen value of the pivot scale, $k_0$.  This is because, when $A_{\rm s}$ is held fixed, $d C_\ell/dn_{\rm s}$ will be small near $k_0$.  Different $k_0$ would correspond to {\em physically} different spectra; our calculations are for the case of $k_0 = 0.05$ Mpc$^{-1}$, which is representative of pivot scales corresponding to the high-$\ell$ region.  Also, defining a gradient for $n_{\rm s}$ might be difficult for scales approaching the Hubble scale today.  However, most of the constraining power comes from much smaller scales, where such ambiguity does not arise.

We now turn to the question of the error on $\Delta X/X_0$ for a {\em single} parameter from an idealized (i.e.\ cosmic variance limited) experiment. This could be studied in more detail using a Fisher matrix formalism, but since it is unclear which set of parameter variations should be considered, we leave this study for the future.

 The correlation matrix $C_{\ell, m, \ell+1, m}$ has $2\ell+1$ off-diagonal modes for each $\ell$, which are diagonal in $m$ for a  dipole orientated along the polar direction. The variance of each  mode is, to lowest order in $\Delta X$, $C_{\ell} C_{\ell+1}$. One can therefore construct the total signal-to-noise of the correlations by summing over $\ell$ and $m$ (see e.g.\ \cite{Amendola:2010ty})
 \beq \label{eqn:sn}
 \left( \frac{S}{N} \right)^2 = \sum_{\ell=2}^{\ell_{\rm max}}
 \sum_{m=-\ell}^{\ell}
 \frac{C_{\ell,m,\ell+1,m}^2}{ C_{\ell}^{\rm }  C_{\ell+1}^{\rm }}\,.
\eeq

 For a general dipole direction, the harmonic modes are rotated according to $a_{\ell m}' = \sum_{m'} D^{\ell}_{m' m} (\phi,\psi,\gamma) a_{\ell m'}$, where $\phi, \psi, \gamma$ are the Euler angles and $D^{\ell}_{m' m}$ the Wigner D-matrix. Since we orient the (unrotated) dipole in the polar direction, $\gamma=0$. The correlations are then redistributed between non-diagonal $m$ modes, depending on the Euler angles. However, one can show that the $S/N$ is invariant under rotations, so~(\ref{eqn:sn}) holds for any direction.

\begin{figure}
\centering
\includegraphics[width=\columnwidth,angle=0]{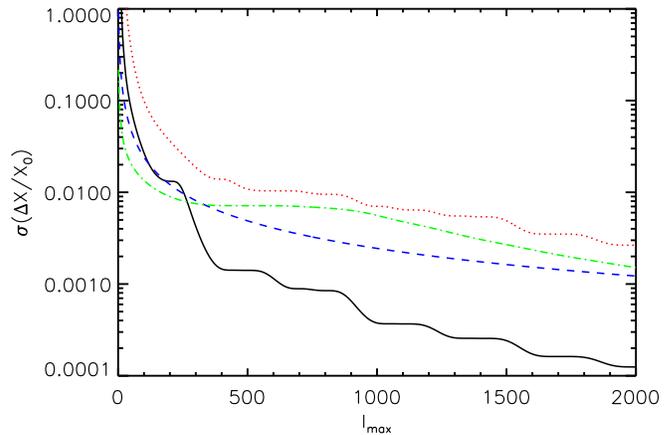} \caption{\label{fig:sigma} Fractional error on $\Delta X/X_0$ for a cosmic variance limited experiment out to $\ell = \ell_{\rm max}$.  Labeling is the same as in Fig.~\ref{fig:cls}.}
\end{figure}

An estimate of the error on $\Delta X/X_0$ can be derived from the total $S/N$ by $\sigma (\Delta X/X_0) =\Delta X /(X_0\, S/N)$~\cite{Amendola:2010ty}. In Fig.~\ref{fig:sigma} we plot this error as a function of $\ell_{\rm max}$. At $\ell_{\rm max} = 2000$, which is roughly the {\em Planck\/} satellite cosmic variance limit, $\sigma (\Delta X/X_0) \approx 10^{-4}$ for $\alpha_{\rm e}$ and $\sim$$10^{-3}$ for the other cosmological parameters. This compares to $\sigma (\beta) \sim 2 \times 10^{-4}$ for the aberration effect~\cite{Kosowsky:2010jm,Amendola:2010ty}.

Returning to the case of $\alpha_{\rm e}$, if we assume that the dipole variation claimed from quasar spectra at $z\simeq 2$  continues as a linear slope in look-back time, then $\Delta \alpha_{\rm e}/\alpha_{\rm e} \approx 1.5 \times 10^{-5}$ at $t_{\rm rec}$. Supposing instead that the gradient is linear in comoving distance $r(z)$, then $\Delta \alpha_{\rm e}/\alpha_{\rm e}$ would be a factor of roughly two higher at the LSS.  However, this is still below the detectable limit from measurements of the CMB according to our results.

In practice, the variation among different parameter gradients may contain partial degeneracies, and hence disentangling various possibilities may be challenging.  Also, for small $\Delta X$, it may be difficult to distinguish the correlations from those due to the aberration effect.  On the other hand, a parameter gradient producing correlations significantly larger than those of aberration should be easy to detect.  Also, weaker gradients may be detectable in the future using the extra information contained in 21 cm surveys \cite{Lewis:2007kz}.

One can also ask what happens to CMB polarization anisotropies in the presence of gradients. This has been studied for the case of a dipole modulation in the 3-dimensional potential (e.g.\ \cite{Dvorkin:2007jp}), which corresponds to our case $X = A_s$, and one finds similar correlations between neighbouring multipoles in polarization.  This additional data can improve sensitivity by a factor of order two for the abberation effect~\cite{Amendola:2010ty}, and we expect the same to hold true for the cosmological parameter variations.

The case $X = A_s$ has recently been investigated in the {\em WMAP} data~\cite{Hanson:2009gu}. The authors found marginal evidence for a dipole modulation with amplitude $\Delta A_{\rm s}/A_{\rm s} = 0.07$ at $\ell < 60$,  but found the effect decreased at higher $\ell$. It would therefore be interesting to check the {\em WMAP} data for other parameters with different correlation spectra than the primordial amplitude.
 

{\em Conclusions.---}%
The spatial variation of fundamental constants has been considered before (e.g., \cite{BarrowOToole01,Donoghue03,Sigurdson03}), but here we have shown how a gradient across our Hubble patch generically gives rise to correlations between neighboring multipoles in the CMB anisotropy spectrum.

Obviously we could extend the dipole to the case of quadrupolar modulation, giving correlations between multipole moments with $\left|\Delta \ell\right| = 2$, and in principle go beyond the quadrupole.  However, the physical interpretation then becomes less clear, and one would need an underlying model to relate all the degrees of freedom.

Returning to the simplest dipole case, if such a ``tilt'' was found in a parameter, then an explanation would have to be sought in the very early Universe.  Inflationary models have been suggested which lead to anisotropic initial power spectra~\cite{Gumrukcuoglu:2007bx,Ackerman:2007nb} and specific models might also explain spatial gradients in parameters.  Other possibilities include modulated reheating (e.g.\ \cite{Bond:2009xx}), bubble collisions (e.g.\ \cite{Chang:2008gj}) and topological defects (e.g.\ \cite{2008MNRAS.390..913C}). 
Certainly it is worth testing future data sets for these correlations, since they could be a smoking gun for variations in physics on the largest accessible scales. 


{\em Acknowledgments.---}%
This research was supported by the Natural Sciences and Engineering Research Council of Canada.  We thank Ted Bunn, Arthur Kosowksy, Ali Narimani, and Kris Sigurdson for useful discussions. 

\bibliography{bib}

\end{document}